\documentclass{aastex}
\usepackage{emulateapj5}
\usepackage{onecolfloat5}
\usepackage{apjfonts}
\usepackage{epsfig}

\newcommand{\be}{\begin{equation}}
\newcommand{\beq}{\begin{equation}}
\newcommand{\ba}{\begin{eqnarray}}
\newcommand{\ee}{\end{equation}}
\newcommand{\eeq}{\end{equation}}
\newcommand{\ea}{\end{eqnarray}}

\newcommand{\hs}{\hspace{1mm}}

\def\lsim{~\rlap{$<$}{\lower 1.0ex\hbox{$\sim$}}}

\def\gsim{~\rlap{$>$}{\lower 1.0ex\hbox{$\sim$}}}

\begin{document}
\twocolumn[
\submitted{Submitted to ApJ}
\title{Continuum Emission by Cooling Clouds}

\author{Mark Dijkstra}
\affil{Harvard-Smithsonian Center for Astrophysics, 60 Garden Street, Cambridge, MA 02138, USA}

\label{firstpage}
\begin{abstract}
The collapse of baryons into the center of a host dark matter halo is
accompanied by radiation that may be detectable as compact ($\lsim 10$ kpc) UV-continuum and Ly$\alpha$ emission with Ly$\alpha$ luminosities as high as $\sim 10^{42}-10^{43}$ erg s$^{-1}$ in halos of mass $M=10^{11}-10^{12}M_{\odot}$. We show that the {\it observed} equivalent width (EW) of the Ly$\alpha$ line emitted by these cooling clouds is EW$\lsim 400$ \AA\hs(restframe). These luminosities and EWs are comparable to those detected in narrowband surveys for
redshifted Ly$\alpha$ emission. The rest-frame ultraviolet of Ly$\alpha$ emitting cooling clouds radiation may be dominated by two-photon transitions from 2s$\rightarrow$1s. The resulting spectrum can distinguish cooling clouds from a broad class of young star forming galaxies.
 \end{abstract}

]
 
\section{Introduction}
\label{sec:intro}

Gas collapse into the center of a host dark matter halo requires the
release of gravitational binding energy in the form of cooling
radiation. The efficiency of gas cooling therefore plays an important
role in the formation of stars and galaxies. For example, it has been
shown that trapping of cooling radiation can affect the fragmentation
of a collapsing gas cloud \citep{SS06}. Understanding cooling
radiation is therefore essential in our understanding of galaxy
formation. A large fraction of cooling radiation may emerge as
Ly$\alpha$ emission \citep{HS00,Fardal01}, possibly accompanied by
emission from helium \citep{Yang06}. Galaxies may therefore be visible
as extended Ly$\alpha$ sources before star formation occurs. This realization coincided with the discovery of two large
Ly$\alpha$ 'blobs' (LABs) by \citet{Steidel00}\footnote{Extended
Ly$\alpha$ emission is quite common around radio galaxies, and is
thought to be powered by some sort of jet-IGM interaction
\citep[e.g.][]{Chambers90,V07}. However, the LABs discovered by
\citet{Steidel00} have no associated radio-galaxies.}. Naturally, LABs
have been associated with cooling radiation.

Since the discovery of LABs by \citet{Steidel00} several tens of new
LABs have been discovered \citep{Matsuda04,Dey05,Saito06,Ouchi08b}. Various
groups have investigated what powers Ly$\alpha$ emission in LABs
\citep[e.g.][]{Matsuda06,Saito07}. Recent studies suggest that the
majority is likely powered by (obscured) stars and/or AGN
\citep[e.g.][]{Basu04,Matsuda07}, or by a superwind
\citep[e.g.][]{Taniguchi00,Mori04}. However, in some LABs it is
possible to rule out all these mechanisms, as each is expected to
produce characteristic metal lines at flux levels that are not
detected. This has led to the conclusion that cooling radiation
 may have been detected
\citep{Nilsson06,Saito07,Smith07}.

Cooling radiation may also manifest itself as compact Ly$\alpha$
emission (rather than spatially extended) with luminosities comparable
to those observed in high-redshift Ly$\alpha$ emitting
galaxies. \citet{HS00} pointed out that $\sim 30\%$ of their total
computed cooling luminosity may originate from the central $\lsim
10\%$ in radius. Furthermore, \citet{Fardal01} and \citet{Yang06}
found that spatially extended cooling gas clouds in 3-D simulations
contain dense compact clumps ($< 10$ kpc, $n_{\rm H,max} \sim 50$
cm$^{-3}$) which dominate the total Ly$\alpha$ cooling emission. Since
it is plausible that star formation takes place inside these densest
clumps, a major challenge lies in determining which process provides the
biggest contribution to the Ly$\alpha$ luminosity emerging from
clumps: cooling or young stars \citep[e.g.][for a more detailed
discussion]{Keres05}.
\begin{figure*}
\vbox{\centerline{\epsfig{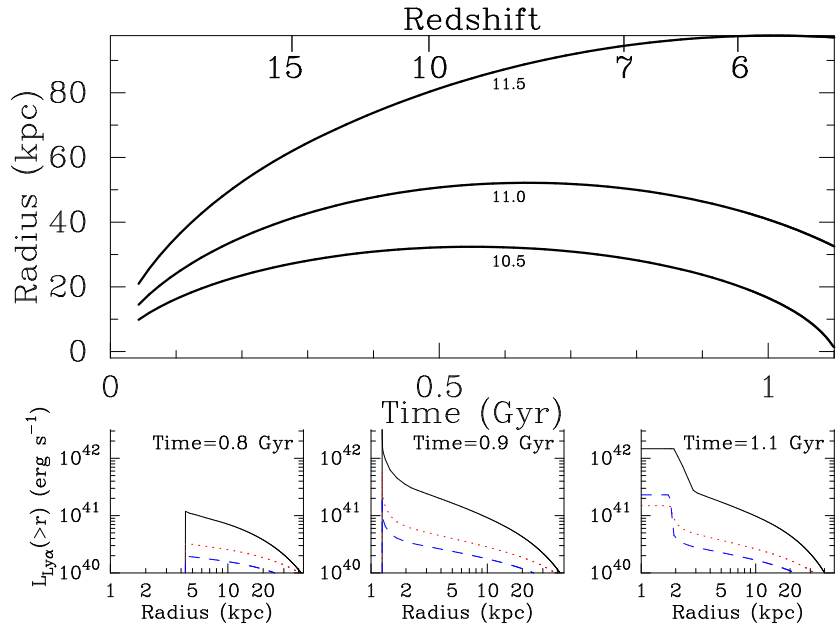}}}
\caption[]{The collapse of a $M_{\rm tot}=10^{11.5}M_{\odot}$ halo in
a 1-D hydrodynamical simulation and its accompanying Ly$\alpha$
cooling radiation are shown. The {\it top panel} shows the radial
evolution of mass shells initially enclosing log$M_{\rm
tot}/M_{\odot}=10.5,11$ and $11.5$ (see labels). The lower horizontal
axis shows the age of the Universe in Gyr, while the upper shows
redshift. The {\it lower three panels} show the total Ly$\alpha$
luminosity {\it outside} a sphere of radius $R$ as a function of $R$
(see text for a justification of this peculiar choice) at three
times. The total Ly$\alpha$ luminosity derived from the 1-D code is
likely underestimated (see text). For comparison, we also show the
luminosity in recombination ({\it dotted line}) and bremsstrahlung
({\it dashed line}) cooling radiation. These processes are clearly
subdominant. In the early stages of the collapse ({\it lower left
panel}) shells at all radii contribute to the total Ly$\alpha$
luminosity. However, during later stages in the collapse the
Ly$\alpha$ luminosity is dominated by emission from the central few
kpc. This Figure illustrates a property of cooling radiation that is
not unique to this spherical collapse model: namely that {\it the
total Ly$\alpha$ luminosity of a cooling cloud may be dominated by
emission from the central few kpc.}}
\label{fig:sim}
\end{figure*}

\citet{D06} quantified the properties of cooling radiation by
performing Monte-Carlo Ly$\alpha$ radiative transfer calculations in
collapsing optically thick gas clouds, and found that Ly$\alpha$ cooling radiation should be systematically blueshifted. In practice it may prove
difficult to actually observe this blueshift, as this requires the
detection of other lines, which are often absent. \citet{D06} also
showed that Ly$\alpha$ cooling radiation may reveal itself through a
frequency dependent surface brightness profile: the surface brightness
profile of the reddest Ly$\alpha$ photons is expected to be flatter
than that of the bluest Ly$\alpha$ photons. However, the strength of
this signal is model dependent and may be too subtle to be
detectable. Therefore, it is desirable to have more diagnostics
that characterize cooling radiation.

The main goal of this paper is to provide new diagnostics to identify
(compact) cooling clouds in data generated by existing and
upcoming telescopes\footnote{ The term 'Ly$\alpha$ cooling radiation' that is used throughout this paper refers to both real (i.e. collisionally excited) cooling radiation, as well as Ly$\alpha$ radiation that is emitted as a by-product of recombination cooling (even though Ly$\alpha$ radiation that is produced in radiative cascades following recombination is technically not cooling radiation, since it does not transport any heat from the gas). In both cases, the Ly$\alpha$ radiation is emitted in a 'proto-galactic' gas cloud that is still collapsing to from stars. Furthermore, for neutral fractions as low as $x_{\rm HI}=10^{-4}$, the 'real' Ly$\alpha$ cooling rate due to collisional excitation, $\Lambda_{{\rm Ly}\alpha}$ is equal to the Ly$\alpha$ emission rate due to recombination (for $T_{\rm gas}=3\times 10^4$ K, see Appendix~\ref{app:reccool}). Hence, even recombining \ion{H}{2} regions can emit a significant amount of Ly$\alpha$ cooling radiation.}.
 We compute the continuum emission of cooling
clouds redward of Ly$\alpha$, and show that this is dominated by
two-photon transitions from 2s$\rightarrow$1s. The spectrum of this
continuum has a peculiar shape which may distinguish cooling
clouds from young star forming galaxies.

The outline of this {\it paper} is as follows. In \S~\ref{sec:sim} we
illustrate that cooling radiation may be detectable as compact
Ly$\alpha$ emission. This is followed by a calculation of the spectrum
of cooling radiation at wavelengths redward of Ly$\alpha$ in
\S~\ref{sec:spectrum}. Finally, in \S~\ref{sec:conc} we discuss the
results of this paper, and present our main conclusions.

\section{Compact Ly$\alpha$ Cooling Radiation}
\label{sec:sim}

We use a modified version of a one--dimensional (spherically
symmetric) hydrodynamical code originally written by \citet{TW95}. For
a detailed description of the code the reader is referred to
\citet{TW95}. The modifications are described in \citet{D04}. The goal
of this section is to emphasis the possibility that the Ly$\alpha$
luminosity of compact cooling clouds is comparable to that of
star forming galaxies.
 
We focus on one particular simulation in which a dark matter halo of
mass $\log M_{\rm tot}/M_{\odot}=11.5$ collapses
at $z_c=4.5$. Such an object is expected to have a Ly$\alpha$ luminosity of $L_{\alpha}\sim$ a few $10^{42}$ erg s$^{-1}$ \citep{HS00}, which is within the range of luminosities that is probed by existing Ly$\alpha$ surveys \citep[e.g.][]{Ouchi08}. At redshift $z=4.5$ the Universe was $\sim
1.4$ Gyr old. We take snap shots of the simulation at intervals of 75
Myr and read out quantities such as gas density, velocity, cooling
rates etc.

The {\it top panel} of Figure~\ref{fig:sim} shows the radial evolution
of mass shells initially enclosing $\log M_{\rm
tot}/M_{\odot}=10.5,11$ and $11.5$ (see labels). The lower horizontal
axis shows the age of the Universe in Gyr, while the upper shows
redshift. Each shell initially expands with the Hubble flow and is
subsequently decelerated by the enclosed mass until it collapses. The
{\it lower three panels} show the total Ly$\alpha$ luminosity {\it
outside} a sphere of radius $R$ as a function of $R$, i.e. $L_{{\rm
Ly}\alpha}(R)=4\pi\int_R^{\infty}r^2drj_{\alpha}(r)$. The reason for
plotting this quantity is that the Ly$\alpha$ volume emissivity,
$j_{\alpha}(r)$, increases rapidly toward smaller radii. This causes
the total Ly$\alpha$ luminosity emitted by material enclosed within
radius $R$ (which is $4\pi\int_{r_{\rm min}}^{R}r^2drj_{\alpha}(r)$)
to depend strongly on the choice of $r_{\rm min}$. Furthermore,
$j_{\alpha}(r)$ is highly uncertain at small radii, because various
 effects that are not included in the simulation (e.g. star formation)
 can affect $j_{\alpha}(r)$, $r_{\rm min}$ and the gas' central density and
velocity field. The Ly$\alpha$ luminosity plotted in
Figure~\ref{fig:sim} is only affected at the smallest radii by these
uncertainties.

The {\it lower left panel} of Figure~\ref{fig:sim} shows that all
radii contribute to the total Ly$\alpha$ luminosity during the early
stages of the collapse. However, at later times the density in the
inner few kpc builds up (e.g. at time=1.1 Gyr, the number density of
hydrogen nuclei is $\gsim 0.01$ cm$^{-3}$ for r$\lsim 4$ kpc, see
Appendix~\ref{app}) which boosts the collisional excitation rate of
the Ly$\alpha$ transition. This causes the {\it central few kpc of the
collapsing gas to dominate the total Ly$\alpha$ cooling luminosity.}

Of course, the results presented above are suspicious as they were
obtained from a spherically symmetric simulation. In a scenario of
hierarchical structure formation halos are build up of mergers of
smaller halos that collapsed at earlier times. Gas accretion is
therefore likely to occur in dense clumps, which affects the results
we obtained above in several ways: (i) Collisions between gas clumps
convert the clumps' kinetic energy (due to infall) into thermal
energy, which is then radiated away as cooling radiation; (ii)
Clumping can boost the the collisional excitation rate of atomic
hydrogen and hence the Ly$\alpha$ emissivity. 

Only (i) affects the predicted surface brightness profile: In the 1-D
calculation, gravitational binding energy is mainly converted into
kinetic energy of the infalling gas. This kinetic energy must be
(partly) released as cooling radiation in the center of the halo
\citep[see][for more detailed discussions on this
topic]{Birnboim03,Keres05}. Therefore, most of the gravitational
binding energy can only be released in the center, which causes
cooling radiation emerging from a 1-D simulation to peak too much
toward the center. However, \citet{HS00} showed that clumpy accretion
also results in surface brightness profile that is peaked towards the
center, and that the inner 10\% of the gas (in radius) emits $\sim
30\%$ of total Ly$\alpha$ luminosity. In the calculations of
\citet{HS00} gravitational binding energy is emitted continuously
while the gas streams towards the center of the halo. Furthermore, 3-D
simulations by \citet{Keres05} have shown that the gas' infall speed
increases down to $\sim 0.5$ virial radii after which the infall
velocity slowly decreases \citep[also see][]{Wise07}. In this case
gravitational potential energy is radiated away in a more compact
region than that described by \citet{HS00}. Therefore, we conclude
that the actual compactness of Ly$\alpha$ cooling radiation is
probably bracketed by the 1-D simulation shown above and the analytic
estimate presented by \citet{HS00}.

The lack of clumps causes the total Ly$\alpha$ luminosities shown in
Figure~\ref{fig:sim} to be lower than the estimates presented by
\citet{HS00}, which only depended only on the energy budget of the
gas. \citet{HS00} found that the total Ly$\alpha$ luminosity due to
cooling can be as large as $10^{43}-10^{44}$ erg s$^{-1}$ in halos in
the mass range $\log M_{\rm tot}/M_{\odot}=11-12$. The compactness of
the emission implies that a cooling blob $\lsim 1$ arcsec in diameter
may emit Ly$\alpha$ luminosities exceeding $10^{43}$ erg
s$^{-1}$. This Ly$\alpha$ luminosity is comparable to that observed
high-redshift Ly$\alpha$ emitting galaxies \citep[e.g.][]{K06}. It is
worth pointing out that resonant scattering of Ly$\alpha$ is not
expected to flatten the observed surface brightness profile
significantly (Dijkstra et al. 2006, hence these source also appear
compact on the sky).

It should be pointed out that in the hierarchical structure formation model, the progenitor halos that merge into the halo of interest, are likely substantially more massive that the minimum halo mass that is capable of cooling gas by collisionally exciting atomic transitions (i.e. halos with $T_{\rm vir} \geq 10^4$ K which corresponds to $M_{\rm tot} \gsim 10^8M_{\odot}$). Hence, gas inside these halos were likely capable of collapsing and forming stars. It is therefore plausible that previously formed stars already exist in the newly formed halo in which gas still needs to cool and collapse to its center. These older stars however are not expected to contribute to the rest-frame UV (and Ly$\alpha$) luminosity of the cloud.

\section{Cooling Radiation Redward of Ly$\alpha$}
\label{sec:spectrum}

In the previous section we concluded that the Ly$\alpha$ luminosity
associated with compact cooling gas clouds can be comparable to that
of star forming galaxies. In this section we calculate the spectral
energy distribution of cooling radiation at wavelengths other than
Ly$\alpha$, and investigate whether this can distinguish a cooling
cloud from a star forming galaxy.

\subsection{The 2-$\gamma$ Continuum: Luminosity \& Spectrum}
\label{sec:2g}

\citet{Yang06} found that gas in their simulations becomes
self-shielding for densities exceeding $n\gsim 200 \bar{n}$, where
$\bar{n}$ is the mean cosmic number density of hydrogen (see their
Fig~2). In the absence of a photoionizing background, collisional
excitation of hydrogen dominates the total cooling rate for gas
temperatures of $\sim 1-3\times 10^4$ K \citep[see e.g Fig~1
of][]{TW95}. For higher temperatures the residual neutral hydrogen
fraction becomes too small ($x_{\rm HI} \lsim 1\times 10^{-3}$
assuming collisional ionization equilibrium, see e.g. House 1964, Hui
\& Gnedin 1997), and this cooling mechanism is shut off. Even in the
1-D simulation used in \S~\ref{sec:sim}, which did not include
self-shielding, collisional excitation of hydrogen dominates the
cooling rate at radii $\lsim 10$ kpc (see Fig~\ref{fig:sim}).

\begin{figure}[t]
\vbox{\centerline{\epsfig{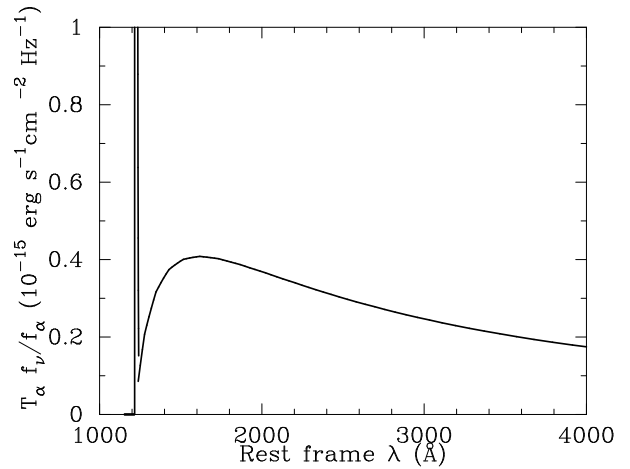}}}
\caption[]{Spectral energy distribution of cooling radiation emitted
by gas at $T\sim 2\times 10^4$ K is shown. A prominent Ly$\alpha$ line
(see Fig~\ref{fig:ew}, the width and height of the line shown here are
schematic.) is accompanied by a continuum redward of Ly$\alpha$. This
continuum is dominated by two-photon transitions from
2s$\rightarrow$1s following collisional excitation of the 2s state of
atomic hydrogen. For comparison, an unreddened starburst is expected
to emit a spectrum of the form $f_{\nu}=$constant. The units of
$f_{\nu}$ are expressed in terms of the observed Ly$\alpha$ line flux
(see text).}
\label{fig:spec}
\end{figure}

Collisions between electrons and neutral hydrogen atoms mainly excite
the $n=2$ states (collisional excitation of the $n=3$ states
occurs $\gsim 10$ times less frequently at the temperatures considered
here). The collisional excitation rates from 1s$\rightarrow$2p and
1s$\rightarrow$2s are comparable as the effective collision strengths
are comparable, $\Omega({\rm 1s,2s})/\Omega({\rm 1s,2p})=0.6-0.5$ for
$T_{\rm gas}=1-2\times 10^4$ K \citep[][Table 3.16, the temperature
dependence of this ratio is weak]{OF06}\footnote{Collisional
deexcitation of the 2s state suppresses the $2-\gamma$ emissivity by a
factor of $(1+n_{\rm p}/n_c)^{-1}$, where $n_c=1.5\times 10^4$
cm$^{-3}$ is the critical density of the $2s\rightarrow 1s$ transition
, and $n_p$ is the number density of protons
\citep{OF06}. Therefore, only gas with densities exceeding $n_c$ is
expected to emit the two-photon continuum at levels less than that
calculated above. These high densities are never reached in the
simulation (see Appendix~\ref{app}) and collisional deexcitation is
not important in the context of this paper.}. Transitions from
2s$\rightarrow$1s may only take place when two photons are emitted
that have a combined energy of 10.2 eV \citep{BT40}. This two-photon
process results in the following spectrum
\begin{equation}
f_{2\gamma }(\nu) d\nu=\frac{\Omega({\rm 1s,2s})}{\Omega({\rm
1s,2p})}\dot{N}_{{\rm
Ly}\alpha}h\nu\phi(\nu/\nu_{\alpha})d\nu/\nu_{\alpha},
\label{eq:2g}
\end{equation} where $\dot{N}_{{\rm Ly}\alpha}$ is the rate at which Ly$\alpha$ photons are emitted, $\phi(\nu/\nu_{\alpha})$ $d\nu/\nu_{\alpha}$ is the probability that 1 photon is emitted in the frequency range $\nu \pm d\nu/2$, which can be found in \citet{SG51}. The spectral energy distribution according to Eq~\ref{eq:2g} is shown in Figure~\ref{fig:spec}. The units of $f_{\nu}$ are chosen such that $\int d\nu f_{2\gamma}(\nu)=[\Omega({\rm 1s,2s})/\Omega({\rm 1s,2p})]f_{\alpha}/\mathcal{T}_{\alpha}$. Here, $f_{\alpha}$ is the observed Ly$\alpha$ flux and, and $\mathcal{T}_{\alpha}$ is the fraction of Ly$\alpha$ photons that is transmitted to the observer through the IGM (see below). For example, Figure~\ref{fig:spec} shows that the peak flux density in the two-photon continuum $f_{2\gamma,{\rm max}}=2\times 10^{-32}f_{\alpha,17}(0.2/\mathcal{T}_{\alpha})$ erg s$^{-1}$ cm$^{-2}$ Hz$^{-1}$, where $f_{\alpha}\equiv f_{\alpha,17}\times 10^{-17}$ erg s$^{-1}$ cm$^{-2}$. Also shown is a schematic representation of the Ly$\alpha$ line (its shape is arbitrary). 

Other process which contribute to wavelengths redward of Ly$\alpha$
include recombination and bremsstrahlung cooling. As already mentioned
however, in cooling gas at $T\lsim 3\times 10^4$ K both of these
processes are negligible compared to collisional excitation of
hydrogen (see \S~\ref{sec:tracer} for a more detailed discussion), and
their contribution to the continuum spectrum can be safely ignored.

\begin{figure}[t]
\vbox{\centerline{\epsfig{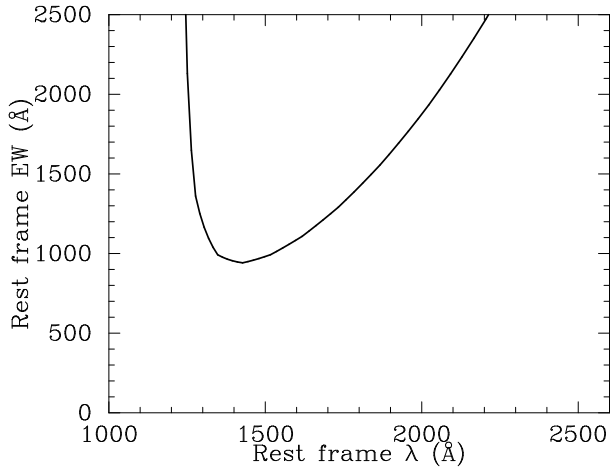}}}
\caption[]{The {\it emitted} rest frame Ly$\alpha$ equivalent width
(EW) of cooling radiation as a function of the wavelength (rest frame)
at which the continuum is measured. The EW diverges when
$\lambda\rightarrow 1216$ \AA, as the continuum flux density
approaches zero. Note that the {\it observed} rest frame EW is likely
to be a factor of $\sim 5-10$ lower because of scattering of
Ly$\alpha$ photons in the IGM (see text).}
\label{fig:ew}
\end{figure}

The prominence of the Ly$\alpha$ line relative to its continuum is
quantified by the equivalent width (EW) which is defined as EW$\equiv
L_{\alpha}/f_{\lambda}$. Because of the strong wavelength dependence
of the continuum between $\lambda=1216-1600$ \AA, the precise EW
depends on the wavelength at which the continuum flux density is
measured. This is shown in Figure~\ref{fig:ew}, which shows the {\it
emitted} rest frame Ly$\alpha$ EW of cooling radiation as a function
of the wavelength (rest frame) at which it is measured. The EW
diverges when $\lambda\rightarrow 1216$ \AA, as the continuum flux
density approaches zero. For $\lambda=1300-1800$ \AA\hs (rest-frame)
the emitted EW is $\sim 1000-1300$ \AA. These EWs are comparable to
those emitted by young metal-poor galaxies \AA\hs \citep{S03}.

The observed rest-frame EW is less than that calculated above as
residual neutral hydrogen in the intergalactic medium (IGM) is
expected to scatter a significant fraction of the emitted Ly$\alpha$
flux out of our line of sight. \citet{D07} calculated that only
$\mathcal{T}_{\alpha}\sim0.1-0.3$ of all Ly$\alpha$ that is emitted by
high-redshift galaxies is actually detected
($\mathcal{T}_{\alpha}=0.2$ was quoted earlier since it corresponds to
the median of this range). Moreover, Ly$\alpha$ that emerges from the
central collapsing clump is likely to be somewhat blueshifted
(\S~\ref{sec:intro}), which results in a preferred lower value of
$\mathcal{T}_{\alpha}$ \citep{D07}. We conclude that the IGM
transmission is very likely $\mathcal{T}_{\alpha}\lsim 0.3$, and that
consequently the {\it observed} rest-frame Ly$\alpha$ EW of cooling
radiation is EW$\lsim 400$ \AA.

The observed rest-frame EW among high redshift Ly$\alpha$ emitting
galaxies can be EW$\sim 100-500$ \AA\hs
\citep[e.g.][]{Dawson04,Shima06}. The existence of these large EW
Ly$\alpha$ emitting galaxies has led to suggestions that primordial
star formation may have been observed
\citep[e.g.][]{MR02,popIII}. Indeed, the results presented above
suggest that some of these galaxies may be powered by cooling
radiation. Figure~\ref{fig:spec} shows that the spectrum of cooling
radiation redward of Ly$\alpha$ peaks around $\lambda\sim 1600$\AA,
with a sharp drop down to $\lambda=1216$\AA. For comparison, a
powerlaw spectrum of the form $f_{\lambda}\propto \lambda^{-2}$ (or
$f_{\nu}=$constant) is expected from an unreddened starburst and
consistent with observations of $27$ candidate $z=6$ galaxies in the
Hubble Ultra Deep Field \citep{Stanway05}. If detected, its continuum
spectrum could therefore distinguish cooling clouds from young
star forming galaxies. This prospect is explored in more detail below.

\subsection{Prospects for Using 2-$\gamma$ Emission as a Tracer of Cooling Clouds}
\label{sec:tracer}

\subsubsection{Cooling Clouds and the 2-$\gamma$ Continuum}

The fate of the gas as it approaches the center of the dark matter
halo is not known \citep[see][for more discussion on
this]{Keres05}. The gas may be slowed down gradually, or may smoothly
join a rotating gaseous disk that formed in the center of the dark
matter halo. In both these scenarios the gas would remain cold and
would emit the spectrum shown in Figure~\ref{fig:spec}. 

However, this is not the case when gas approaches $r=0$ with velocities of order $\sim 100$ km s$^{-1}$ (as was the case in the 1-D simulation used in
\S~\ref{sec:sim}). When the gas crashes into the center, it is
shock-heated to $T\sim 10^6$ K and emits bremsstrahlung up to energies
of $\sim 100$ eV. This ionizing radiation creates an \ion{H}{2} region of
$\lsim$ a few kpc because of the large gas densities in the center of
the halo \citep{Birnboim03}. This ionizing radiation could furthermore doubly ionize helium.

This recombining gas may emit a spectrum that is very
similar to that shown in Figure~\ref{fig:spec}. In
Figure~\ref{fig:hii} the {\it solid line} shows the spectrum emitted
by gas assuming case-A recombination and a gas temperature of
T$=2\times 10^4$ K. In this case the bound-free (free-free) emissivity
is given by $\epsilon=5.3\hs(2.6)\times 10^{-25}$ erg s$^{-1}$
cm$^{3}$ \citep[e.g.][]{OF06}. Furthermore, the total emissivity in
the two-photon continuum is $\epsilon\sim 5.0 \times 10^{-25}$ erg
s$^{-1}$ cm$^{3}$ \citep[][Table 4.1, and we used that
$\epsilon_{2\gamma}\sim0.5\epsilon_{{\rm Ly}\alpha}$]{OF06}. It is
noted that when helium is doubly ionized inside the \ion{H}{2} region, and/or
for higher gas temperatures, more continuum emission is emitted at
$\lambda <2000$ \AA\hs (as well as a \ion{He}{2} H$\alpha$ line at $\lambda=1640$ \AA), which would weaken the appearance of the dip. The
{\it dotted line} shows the spectrum resulting from bound-bound
transitions only (which corresponds exactly to the case shown in
Fig~\ref{fig:spec}). Technically the Ly$\alpha$ + 2-$\gamma$ radiation is not cooling radiation itself, but is emitted by the cooling cloud as a by-product of recombination cooling radiation (see \S~\ref{sec:intro}). Clearly, recombining gas at $T\lsim 2\times 10^4$ K emits a spectrum of similar to that of 'pure' cooling radiation, especially at wavelengths where the spectrum differs from that emitted by star forming galaxies, $\lambda=1200-2000$ \AA.

\subsubsection{Star Forming Galaxies and the 2-$\gamma$ Continuum}
\begin{figure}[t]
\vbox{\centerline{\epsfig{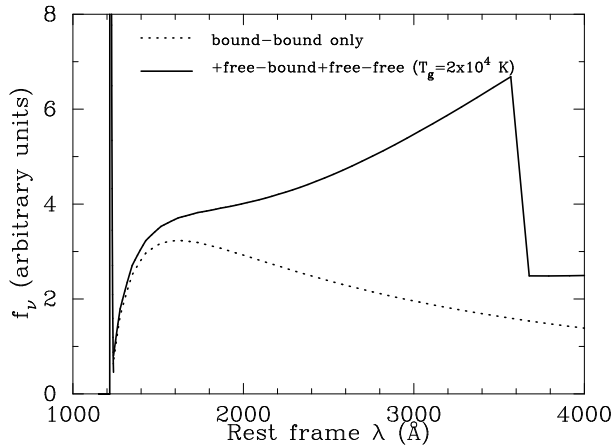}}}
\caption[]{The spectrum emitted by recombining gas at T$=2\times 10^4$
K is shown ({\it solid line}). This spectrum includes the contribution
from free-free, free-bound and bound-bound transitions. The {\it
dotted line} shows the contribution from bound-bound transitions only,
and this spectrum is identical to that of cooling gas
(Fig~\ref{fig:spec}). Clearly, the spectra of recombining and cooling
gas are very similar at $\lambda\lsim 2000$ \AA.}
\label{fig:hii}
\end{figure}

Of course, recombining \ion{H}{2} regions (of primordial composition)
surrounding stars or quasars emit the same spectrum as that shown in Fig~\ref{fig:hii}. However, in observed galaxies this nebular emission is subdominant to the continuum radiation from stars/quasars, which does not exhibit the dip
shown in Figure~\ref{fig:hii}. A notable exception is discussed by
Schaerer (2002), who shows that nebular emission dominates the stellar
continuum at wavelengths redward of Ly$\alpha$ during the first 1-2
Myr in metal-free galaxies that are forming stars according to
top-heavy initial mass function (IMF). Hence, these primordial
star forming galaxies also exhibit the two-photon dip in their spectra
during the first $1-2$ Myr of their lifetime (see Fig 5 of Schaerer
2002). On the other hand, these galaxies also emit \ion{He}{3} recombination lines such as the He 1640 line, which is not the case for cooling radiation when the gas does not get shock heated to $T\gsim 10^5$ K (as was discussed above in \S~3.2.1).

\subsubsection{The Impact of Ly$\alpha$ Radiative Transfer}

It is worth pointing out is that for sufficiently large
columns of neutral hydrogen gas ($N_{\rm HI}\gsim 10^{22}$ cm$^{-2}$),
Ly$\alpha$ photons scatter so frequently that collisional deexcitation
from the 2p level may become important, and the Ly$\alpha$ cooling
channel may be suppressed strongly. This process has been studied in
more detail by \citet{SS06}. However, this mechanism only becomes
efficient when densities exceed $\gsim 10^2$ cm$^{-3}$, which is much
larger than the highest densities in the simulation shown in
Figure~\ref{fig:sim} (see Fig~\ref{fig:app} in
Appendix~\ref{fig:app}). Furthermore, this effect enhances the
expected 2-$\gamma$ continuum relative to the Ly$\alpha$ line, and if
Ly$\alpha$ were ever detected from these ultracompact cooling cores,
then the associated 2-$\gamma$ flux is relatively brighter and thus
easier to detect.

So far, the role of dust in the Ly$\alpha$ radiative transfer has not been discussed. One might expect little dust to exist in gas that is still cooling to form stars. However, as mentioned in \S~\ref{sec:sim} in models of hierarchical structure formation, star formation is expected to have occurred in the less massive progenitor halos which merged into the halo of interest. Thus, dust is expected to exist in the cooling cloud. Dust would redden and suppress the UV-continuum, but to quantify this effect theoretically is difficult. Empirically, \citet{Bouwens06} found the rest-frame UV continuum of $z=6$ Lyman Break Galaxies (LBGs) to be suppressed by dust by a factor of only $\sim 1.4$. It is reasonably to expect the progenitor halos not to contain more dust than these more massive -and hence probably more evolved- LBGs, and we therefore do not expect dust attenuation of continuum cooling radiation to be significant\footnote{Gnedin et al. (2008) found the mean escape fraction of {\it ionizing} radiation in (adaptive mesh refinement) hydrodynamical simulations of high-redshift galaxies to be $\langle f_{\rm esc}\rangle \sim 1\%$. However, the opacity to ionizing photons in these simulations was found to be dominated by neutral atomic hydrogen and helium.}.
 
The impact of dust on Ly$\alpha$ is less clear. In principle, resonant scattering enhances the total distance Ly$\alpha$ photons travel through a medium, which makes them more prone to destruction by dust. In this case, the 2-$\gamma$ continuum would be enhanced relative to Ly$\alpha$ and would therefore be easier to detect. On the other hand, Ly$\alpha$ may preferentially avoid destruction by dust in a two-phase medium in which dust resides in cold ($T\sim 10^4$ K) neutral denser clumps embedded in a hot medium \citep[][]{Neufeld91,Hansen06}. In this scenario, the 2-$\gamma$ continuum would be attenuated (by a factor of $\lsim 1.4$ as we argued above), but the Ly$\alpha$ would be freely transmitted. It is therefore good to keep in mind that our quoted Ly$\alpha$ luminosities have not corrected for dust attenuation, but that this correction is not necessarily needed.

\section{Discussion \& Conclusions}
\label{sec:conc}

The collapse and condensation of baryons into the center of their host
dark matter halos is accompanied by emission of radiation that may be
detectable as compact UV-continuum and Ly$\alpha$ 
emission with Ly$\alpha$ luminosities as high as $\sim
10^{42}-10^{43}$ erg s$^{-1}$ in halos of mass
$M=10^{11}-10^{12}M_{\odot}$. These luminosities are comparable to
those detected in narrowband surveys for redshifted Ly$\alpha$
emission.

Cooling clouds emit continuum radiation redward of Ly$\alpha$ that is dominated by two-photon transitions from
2s$\rightarrow$1s. The {\it emitted} Ly$\alpha$ equivalent width (EW) relative to this continuum is EW$\sim 1000-1300$ \AA\hs(restframe), which is
comparable to the EW emitted by galaxies forming either metal-poor or
very massive stars. The IGM is likely to transmit only $\lsim 30\%$ of
this Ly$\alpha$ flux and the {\it observed} rest frame equivalent
width is therefore EW$\lsim 400$ \AA. This is comparable to the
observed equivalent widths among high-redshift Ly$\alpha$ emitting
galaxies. It is therefore possible that existing surveys have detected
compact Ly$\alpha$ emission that was emitted by cooling clouds.

Because of the dominance of the two-photon continuum, cooling clouds
emit a spectrum that peaks at $\lambda\sim 1600$\AA\hs(rest-frame),
and sharply drops down to $\lambda=1216$\AA. This peculiar spectral
shape distinguishes cooling clouds from observed young star forming
galaxies, whose continuum is dominated by radiation from stars/quasars
\citep[e.g.][]{Stanway05}. However, this is not true for metal-free
galaxies that are forming stars according to top-heavy initial mass
function. In these primordial galaxies, nebular emission may dominate
the continuum redward of Ly$\alpha$ during the first few Myr of their
lifetime (Schaerer 2002). On the other hand, these galaxies also emit
 \ion{He}{3} recombination lines such as the He 1640 line, which is not the case
for cooling radiation when the gas is not shock heated to $T\gsim 10^4$ K (see \S~\ref{sec:tracer}). Furthermore, deeper Ly$\alpha$ observations of compact cooling clouds should find weaker emission on larger scales. However, the same
applies to Ly$\alpha$ emitting star forming galaxies, which are likely
surrounded by faint Ly$\alpha$ halos due to resonant scattering in the
surrounding IGM \citep[e.g.][]{LR99,DL08}. This scattered Ly$\alpha$
emission is not accompanied by continuum radiation redward of the
Ly$\alpha$ line, which could distinguish it from
spatially extended Ly$\alpha$ cooling radiation (polarimetry may
provide additional insights, see Dijkstra \& Loeb 2008). 

Ongoing and future surveys for redshifted Ly$\alpha$ emission aim to
detect Ly$\alpha$ emitters out to redshifts as high as $z\sim 6-12$
\citep[e.g.][and references therein]{StarkLF,Nilsson07}. These
observations will provide new insights into the epoch of reionization
and galaxy formation at high redshift. Primordial galaxies -like
cooling clouds- may emit very large EW Ly$\alpha$ emission lines
\citep[e.g.][]{S03}. In order to measure the EW of the Ly$\alpha$ line
and to determine whether primordial galaxies are detected, the
continuum redward of Ly$\alpha$ must be detected. This is challenging;
if the observed flux in the Ly$\alpha$ line is $f_{\alpha,17}$ (in
units of $10^{-17}$ erg s$^{-1}$ cm$^{-2}$) and its equivalent width
is EW (in \AA), then the AB-magnitude of the continuum is $m_{\rm
AB}=30.4-2.5\log f_{\alpha,17}+2.5\log ({\rm EW}/200 {\rm \AA})$. For
comparison, the Hubble Ultra Deep Field reached an $8-\sigma$ limiting
magnitude in the z' filter of $z'_{\rm AB}=28.5$ \citep{Bunker04}.

Detecting both the Ly$\alpha$ line and the continuum is not only
desirable for identifying primordial galaxies. An additional advantage
of combining Ly$\alpha$ and continuum measurements of high-redshift
galaxies is that it can greatly improve the constraints on the epoch
of reionization \citep[e.g.][]{K06,McQuinn07} over constraints
obtained using Ly$\alpha$ alone. The work presented in this paper
suggests that if primordial galaxies can be detected and identified,
then the same applies to cooling clouds. Moreover, primordial
galaxies may only be luminous in Ly$\alpha$ for $\lsim 10$ Myr
\citep{MR02,popIII}. Cooling clouds could be luminous in Ly$\alpha$
for longer than 10 Myr, which would make them more easily detectable
than primordial galaxies. The discovery of two-photon dominated
Ly$\alpha$ emitters would offer us an exciting glimpse of either the process of gas cooling in newly forming galaxies, or of star formation in the first primordial galaxies.

{\bf Acknowledgments} I thank Marco Spaans for his comments on an
earlier draft of this paper, and Zolt\'{a}n Haiman for useful
discussions. This work was supported by Harvard University funds.

\appendix
\section{Output from the Simulation}

\begin{figure}[t]
\vbox{\centerline{\epsfig{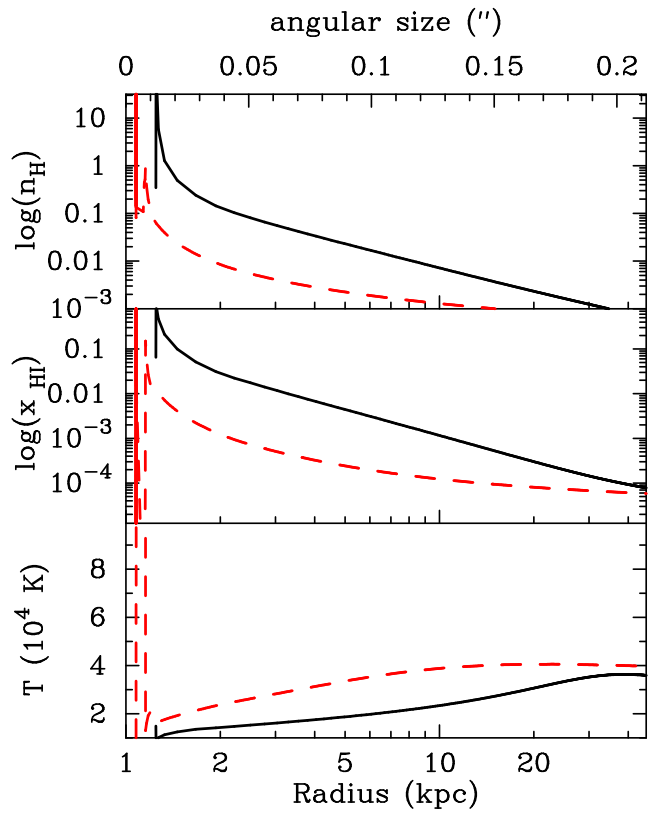}}}
\caption[]{Simulation outputs at time=0.9 Gyr ({\it solid lines}) and
time=1.1 Gyr ({\it dashed lines}) are shown. The {\it
upper/middel/lower panel} shows the radial dependence of the gas
density/ neutral fraction of atomic hydrogen/temperature. These
quantities were used to compute the cooling rates shown in
Figure~\ref{fig:sim}. The snapshot at time=1.1 Gyr shows that the
simulation behaves strangely in the inner 2 kpc (see text).}
\label{fig:app}
\end{figure}

Two snapshots of the simulation used in \S~\ref{sec:sim} are shown in
Figure~\ref{fig:app}. The radial dependence of the number density of
H-nuclei ($n_{\rm H}$, {\it upper panel}), the fraction of neutral
atomic hydrogen gas ($x_{\rm HI}=n_{\rm HI}/n_{\rm H}$, {\it central
panel}), and the gas temperature (T, {\it lower panels}) are shown at
time=0.9 Gyr ({\it solid lines}) and time=1.1 Gyr ({\it dashed
lines}). Snapshots of the simulation at these times were also shown in
Figure~\ref{fig:sim}.

\label{app}

Figure~\ref{fig:app} shows that the gas temperature in the inner 10
kpc is $T_{\rm gas}\lsim 3 \times 10^4$ K at both times, which corresponds to the temperature regime where collisional excitation of hydrogen dominates
gas cooling. Note that the snapshot at time=1.1 Gyr shows some strange behaviour at $r\lsim 2$ kpc. This is related to the formation of a shock that is produced by the gas that is radially moving inwards at a supersonic speed, which causes the gas to shock-heat to $T> 10^5$ K. As was mentioned in \S~\ref{sec:sim}, the simulation results are suspicious in these innermost regions.

Lastly, we point out that the gas density is well below the critical
density, $n_c=1.5 \times 10^4$ cm$^{-3}$, at which collisional
deexcitation of the 2s-state becomes important. 

\section{Ly$\alpha$ Recombination vs. Ly$\alpha$ Cooling Radiation}
\label{app:reccool}
Here we compare the Ly$\alpha$ cooling rate due to collisional excitation, $\Lambda_{{\rm Ly}\alpha}$, with the Ly$\alpha$ emission rate due to recombination, $R_{{\rm Ly}\alpha}$. Their ratio is given by

\begin{equation}
\mathcal{R}\equiv\frac{\Lambda_{{\rm Ly}\alpha}}{R_{{\rm Ly}\alpha}}=\frac{7.3\times 10^{-19}x_{\rm HI}\exp\big{(}\frac{-118400}{T}\big{)}}{0.68\alpha_{\rm rec,B}\times h\nu_{\alpha}\times (1.0-x_{\rm HI})},
\end{equation} where $\alpha_{\rm rec,B}=2.6 \times 10^{-13}(T/10^4\hs{\rm K})^{-0.7}$ cm$^3$ s$^{-1}$ is the case-B recombination coefficient. For $T=3\times 10^4$ K we find $\mathcal{R}=1.0\times (x_{\rm HI}/10^{-4})$ (for $x_{\rm HI}\ll 1$), and $\mathcal{R}$ increases with $T$. That is, even in a highly ionized plasma, a significant fraction of the emitted Ly$\alpha$ photons are produced following collisional excitation of the much rarer neutral hydrogen atoms.

\label{lastpage}

\begin{thebibliography}{xx}
\expandafter\ifx\csname natexlab\endcsname\relax\def\natexlab#1{#1}\fi

\bibitem[Basu-Zych \& Scharf(2004)]{Basu04} Basu-Zych, A., \&  Scharf,
C.\ 2004, \apjl, 615, L85

\bibitem[Birnboim \& Dekel(2003)]{Birnboim03} Birnboim, Y., \&  Dekel,
A.\ 2003, \mnras, 345, 349

\bibitem[Bouwens et al.(2006)]{Bouwens06} Bouwens, R.~J., 
Illingworth, G.~D., Blakeslee, J.~P., \& Franx, M.\ 2006, \apj, 653, 53 

\bibitem[Breit \& Teller(1940)]{BT40} Breit, G., \& Teller,  E.\ 1940,
\apj, 91, 215

\bibitem[Bunker et al.(2004)]{Bunker04} Bunker, A.~J., Stanway,
E.~R., Ellis, R.~S., \& McMahon, R.~G.\ 2004, \mnras, 355, 374

\bibitem[Chambers et al.(1990)]{Chambers90} Chambers, K.~C.,  Miley,
G.~K., \& van Breugel, W.~J.~M.\ 1990, \apj, 363, 21

\bibitem[Dawson et al.(2004)]{Dawson04} Dawson, S., et al.\  2004,
\apj, 617, 707

\bibitem[Dey et al.(2005)]{Dey05} Dey, A., et al.\ 2005,  \apj, 629,
654

\bibitem[Dijkstra et al.(2004)]{D04} Dijkstra, M., Haiman,  Z., Rees,
M.~J., \& Weinberg, D.~H.\ 2004, \apj, 601, 666

\bibitem[Dijkstra et al.(2006)]{D06} Dijkstra, M., Haiman,  Z., \&
Spaans, M.\ 2006, \apj, 649, 14

\bibitem[Dijkstra et al.(2007)]{D07} Dijkstra, M., Lidz,  A., \&
Wyithe, J.~S.~B.\ 2007, \mnras, 377, 1175

\bibitem[Dijkstra \& Wyithe(2007)]{popIII} Dijkstra, M., \&  Wyithe,
J.~S.~B.\ 2007, \mnras, 379, 1589
\bibitem[Dijkstra \& Loeb(2008)]{DL08} Dijkstra, M., \& Loeb, A.\ 2008, \mnras, 386, 492 
\bibitem[Fardal et al.(2001)]{Fardal01} Fardal, M.~A., Katz, N.,
Gardner, J.~P., Hernquist, L., Weinberg, D.~H., \& Dav{\'e}, R.\ 2001,
\apj, 562, 605

\bibitem[Gnedin et al.(2008)]{Gnedin08} Gnedin, N.~Y., Kravtsov, 
A.~V., \& Chen, H.-W.\ 2008, \apj, 672, 765 

\bibitem[Haiman et al.(2000)]{HS00} Haiman, Z., Spaans, M.,  \&
Quataert, E.\ 2000, \apjl, 537, L5

\bibitem[Hansen \& Oh(2006)]{Hansen06} Hansen, M., \& Oh, S.~P.\ 2006, \mnras, 367, 979 

\bibitem[House(1964)]{House64} House, L.~L.\ 1964, \apjs, 8,  307

\bibitem[Hui \& Gnedin(1997)]{Hui97} Hui, L., \& Gnedin,  N.~Y.\ 1997,
\mnras, 292, 27

\bibitem[Kashikawa et al.(2006)]{K06} Kashikawa, N., et  al.\ 2006,
\apj, 648, 7

\bibitem[Kere{\v s} et al.(2005)]{Keres05} Kere{\v s}, D.,  Katz, N.,
Weinberg, D.~H., \& Dav{\'e}, R.\ 2005, \mnras, 363, 2

\bibitem[Loeb \& Rybicki(1999)]{LR99} Loeb, A., \& Rybicki,  G.~B.\
1999, \apj, 524, 527

\bibitem[Malhotra \& Rhoads(2002)]{MR02} Malhotra, S., \&  Rhoads,
J.~E.\ 2002, \apjl, 565, L71

\bibitem[Matsuda et al.(2004)]{Matsuda04} Matsuda, Y., et al.\  2004,
\aj, 128, 569

\bibitem[Matsuda et al.(2006)]{Matsuda06} Matsuda, Y., Yamada,  T.,
Hayashino, T., Yamauchi, R., \& Nakamura, Y.\ 2006, \apjl, 640, L123

\bibitem[Matsuda et al.(2007)]{Matsuda07} Matsuda, Y., Iono, D., 
Ohta, K., Yamada, T., Kawabe, R., Hayashino, T., Peck, A.~B., \& Petitpas, 
G.~R.\ 2007, \apj, 667, 667 

\bibitem[McQuinn et al.(2007)]{McQuinn07} McQuinn, M., Lidz, A.,
Zahn, O., Dutta, S., Hernquist, L., \& Zaldarriaga, M.\ 2007, \mnras,
377,  1043

\bibitem[Mori et al.(2004)]{Mori04} Mori, M., Umemura, M., \&
Ferrara, A.\ 2004, \apjl, 613, L97

\bibitem[Neufeld(1991)]{Neufeld91} Neufeld, D.~A.\ 1991, \apjl, 
370, L85 

\bibitem[Nilsson et al.(2006)]{Nilsson06} Nilsson, K.~K., Fynbo,
J.~P.~U., M{\o}ller, P., Sommer-Larsen, J., \& Ledoux, C.\ 2006, \aap,
452,  L23

\bibitem[Nilsson et al.(2007)]{Nilsson07} Nilsson, K.~K., Orsi,  A.,
Lacey, C.~G., Baugh, C.~M., \& Thommes, E.\ 2007, \aap, 474, 385

\bibitem[Osterbrock \& Ferland(2006)]{OF06} Osterbrock,  D.~E., \&
Ferland, G.~J.\ 2006, {\it Astrophysics of gaseous nebulae and active
galactic nuclei}, 2nd.~ed.~by D.E.~Osterbrock and
G.J.~Ferland.~Sausalito, CA: University Science Books, 2006,

\bibitem[Ouchi et al.(2008a)]{Ouchi08} Ouchi, M., et al.\ 2008a, 
\apjs, 176, 301 

\bibitem[Ouchi et al.(2008b)]{Ouchi08b} Ouchi, M., et al.\ 2008b, 
ArXiv e-prints, 807, arXiv:0807.4174 

\bibitem[Saito et al.(2006)]{Saito06} Saito, T., Shimasaku, K.,
Okamura, S., Ouchi, M., Akiyama, M., \& Yoshida, M.\ 2006, \apj, 648,
54

\bibitem[Saito et al.(2007)]{Saito07} Saito, T., Shimasaku, K.,
Okamura, S., Ouchi, M., Akiyama, M., Yoshida, M., \& Ueda, Y.\ 2007,
ArXiv  e-prints, 705, arXiv:0705.1494

\bibitem[Schaerer(2002)]{S02} Schaerer, D.\ 2002, \aap,  382, 28

\bibitem[Schaerer(2003)]{S03} Schaerer, D.\ 2003, \aap,  397, 527

\bibitem[Shimasaku et al.(2006)]{Shima06} Shimasaku, K., et  al.\
2006, \pasj, 58, 313

\bibitem[Smith \& Jarvis(2007)]{Smith07} Smith, D.~J.~B., \&  Jarvis,
M.~J.\ 2007, \mnras, 378, L49

\bibitem[Spaans \& Silk(2006)]{SS06} Spaans, M., \& Silk,  J.\ 2006,
\apj, 652, 902

\bibitem[Spergel et al.(2007)]{Spergel07} Spergel, D.~N., et al.\
2007, \apjs, 170, 377

\bibitem[Spitzer \& Greenstein(1951)]{SG51} Spitzer, L.~J.,  \&
Greenstein, J.~L.\ 1951, \apj, 114, 407

\bibitem[Stanway et al.(2005)]{Stanway05} Stanway, E.~R.,  McMahon,
R.~G., \& Bunker, A.~J.\ 2005, \mnras, 359, 1184

\bibitem[Stark et al.(2007)]{Starklens} Stark, D.~P., Ellis,  R.~S.,
Richard, J., Kneib, J.-P., Smith, G.~P., \& Santos, M.~R.\ 2007,
\apj, 663, 10

\bibitem[Stark et al.(2007)]{StarkLF} Stark, D.~P., Loeb, A., 
\& Ellis, R.~S.\ 2007, \apj, 668, 627 

\bibitem[Steidel et al.(2000)]{Steidel00} Steidel, C.~C.,  Adelberger,
K.~L., Shapley, A.~E., Pettini, M., Dickinson, M., \&  Giavalisco, M.\
2000, \apj, 532, 170

\bibitem[Taniguchi \& Shioya(2000)]{Taniguchi00} Taniguchi, Y., \&
Shioya, Y.\ 2000, \apjl, 532, L13

\bibitem[Thoul \& Weinberg(1995)]{TW95} Thoul, A.~A.~\&  Weinberg,
D.~H.\ 1995, \apj, 442, 480

\bibitem[Villar-Mart{\'{\i}}n(2007)]{V07}  Villar-Mart{\'{\i}}n, M.\
2007, New Astronomy Review, 51, 194

\bibitem[Wise \& Abel(2007)]{Wise07} Wise, J.~H., \& Abel, T.\  2007,
\apj, 665, 899

\bibitem[Yang et al.(2006)]{Yang06} Yang, Y., Zabludoff,  A.~I.,
Dav{\'e}, R., Eisenstein, D.~J., Pinto, P.~A., Katz, N., Weinberg,
D.~H., \& Barton, E.~J.\ 2006, \apj, 640, 539
\end{thebibliography}
\end{document}